\documentclass[pdflatex,sn-mathphys-num]{sn-jnl}

\usepackage{graphicx}%
\usepackage{mathrsfs}
\usepackage{multirow}%
\usepackage{amsmath,amssymb,amsfonts}%
\usepackage{amsthm}%
\usepackage[title]{appendix}%
\usepackage{xcolor}%
\usepackage{longtable}
\usepackage{array}
\usepackage{textcomp}%
\usepackage{manyfoot}%
\usepackage{booktabs}%
\usepackage{algorithm}%
\usepackage{algorithmicx}%
\usepackage{algpseudocode}%
\usepackage{listings}%

\theoremstyle{thmstyleone}%
\theoremstyle{thmstyletwo}%

\theoremstyle{thmstylethree}%

\begin{document}

\title[Article Title]{Decadal Analysis of Delhi’s Air Pollution Crisis: Unraveling the Contributors}


\author*[1]{\fnm{Prachi} \sur{Tewari}}\email{prachii.tewarii@gmail.com}

\author[1]{\fnm{Dr. Shweta} \sur{Jindal}}\email{miss.shweta.singhal@gmail.com}

\affil*[1]{\orgdiv{Department of Information Technology}, \orgname{Indira Gandhi Delhi Technical University for Women}, \orgaddress{\street{Kashmere Gate}, \city{New Delhi}, \postcode{110006}, \state{Delhi}, \country{India}}}

\equalcont{These authors contributed equally to this work.}


\abstract{Recently Delhi has become a chamber of bad air quality. This study explores the trends of probable contributors to Delhi’s deteriorating air quality by analyzing data from 2014 to 2024 -- a period that has not been the central focus of previous research. The study aims to reassess the contributors in light of recent shifts. The consistently worsening air quality has forced the people of Delhi to adapt to an unhealthy environment. People breathing this polluted air are at great risk of developing several health issues such as respiratory infections, heart disease, and lung cancer. The study provides a quantified perspective on how each contributor has influenced pollution levels by identifying percentage contributions of major sources. Over the years, Delhi’s air pollution has been primarily attributed to stubble burning. However, the present study discusses the decline in stubble burning cases in the current scenario and the evolving impact of contributors such as vehicular emissions, industrial activities, and population growth. Moreover, the study assesses the effectiveness of mitigation strategies like Electric Vehicles (EVs), public transport expansion, and pollution control policies. The average levels of the Air Quality Index (AQI) during October-November and November-December remained consistently high from 2018 to 2024, reaching 374 in November 2024. Based on the data-driven analysis, the study demonstrates that existing measures have fallen short and makes a strong case for implementing new long-term  strategies focusing on the root causes.}




\maketitle

\section{Introduction}\label{sec1}

\begin{figure}[H]
    \centering
    \includegraphics[width=0.5\linewidth]{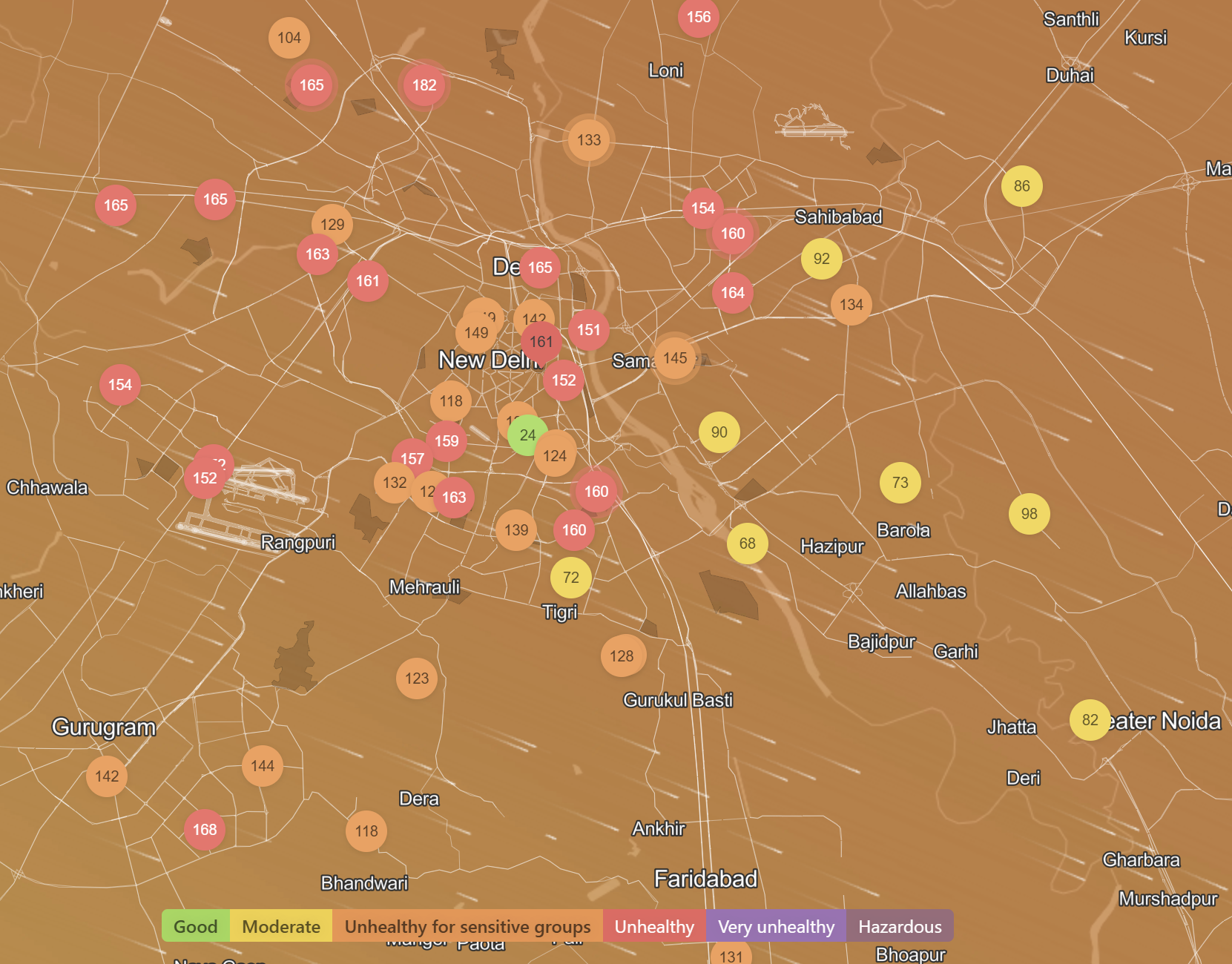}
    \caption{\centering Air Quality Map of Delhi \cite{IQAir2025}}
    \label{fig:air quality map}

\end{figure}

Delhi is a wake-up call to the world on air pollution. Air pollution is the contamination of the indoor or outdoor air by a range of gases and solids that modify its natural characteristics. 
\autoref{fig:air quality map} shows the air quality map of Delhi.
The main pollutants that are harmful to health include particulate matter (PM$_{2.5}$ and PM$_{10}$), Carbon Monoxide (CO), Ozone (O$_3$), Black Carbon (BC), Sulfur Dioxide (SO$_2$) and Nitrogen Oxides (NO$_x$)\cite{WHO2019}. Delhi's air quality was 35 times higher than the safe limit set by the World Health Organization (WHO) \cite{BBC}. According to the WHO, Delhi is the fourth most polluted city in the world in terms of Suspended Particulate Matter (SPM). In terms of air pollution in Delhi, the United Nations Children's Fund (UNICEF) said that an estimated 4.41 million children in the city missed three days of school, following a decision to close its 5,798 schools to minimize the risk of children being exposed to polluted air \cite{UN2016}. Delhi lies in the north of India between latitudes of 28-24-17 and 28-53-00 north and longitudes of 76-50-24 and 77-20-37 east. Delhi is surrounded by the states of Uttar Pradesh and Haryana. 

Delhi has an area of 1,483 km$^2$. Its maximum length is 51.90 km and maximum width is 48.48 km. The average annual rainfall in Delhi is 714 mm, three-fourths of which falls in July, August, and September. Heavy rain in the Yamuna catchment area can result in a dangerous flood situation in the city. During the summer months of April, May, and June, temperatures can rise to 40-45 degrees Celsius; winters can be cold, with temperatures during December and January falling to 4 to 5 degrees Celsius. February, March, October, and November are the climatically best months \cite{DelhiGov2000} . 

Taking into account the available data and trends, the key sources of Delhi air pollution are debatable. Many studies and public discourse attribute the poor air quality of the city to stubble burning in neighboring states during the post-harvest season, compounded by firecracker emissions during the Diwali festival.

Overall, this research provides a framework for understanding the dynamics of air pollution in Delhi, by analyzing multi-year trends and patterns. Highlights gaps in existing narratives by examining a range of contributing factors – from stubble burning and vehicular emissions to industrial activities and population growth. 

\section{Literature Review}\label{sec2}

\begin{scriptsize}
\begin{longtable}{p{4cm} c p{8cm} c}
    \toprule
    \textbf{Title} & \textbf{Year} & \textbf{Finding} &Reference \\
    \midrule
    \endfirsthead

    \toprule
    \textbf{Title} & \textbf{Year} & \textbf{Finding} &Reference \\
    \midrule
    \endhead

    \midrule
    \multicolumn{4}{r}{\textit{Continued on next page}} \\
    \midrule
    \endfoot

    \bottomrule
    \endlastfoot

 Assessment of traffic-generated gaseous and particulate matter emissions and trends over Delhi (2000–2010) & 2014 & The study showed that CO emissions increased by 77\%, NO$_x$ by 29\% (2000–2010) . Two-wheelers contributed most to CO, VOC emissions. Diesel vehicles were major sources of PM$_{10}$, BC, OC. & \cite{Sindhwani2014}
\\
 Statistical assessment of respirable and coarser size ambient aerosol sources and their timeline trend profile determination: A four year study from Delhi & 2015 & The study analyzed PM$_{10}$ and coarser particulate matter (CPM) sources in Delhi (2006–2009) using Positive Matrix Factorization (PMF) and Ensemble Empirical Mode Decomposition (EEMD). &\cite{Yadav2015}
\\
\\        
 Air pollution trends over Indian megacities and their local-to-global implications & 2016 & Delhi had the highest PM$_{10}$ levels, primarily from vehicular and industrial emissions. &\cite{Gurjar2016}
\\
\\ 
Predicting trends in air pollution in Delhi using data mining&2016&The study uses data mining to analyze the existing trends in air pollution in Delhi and make prediction about the future.&\cite{Taneja2016}
\\
\\
 A Study on Effects of Weather, Vehicular Traffic and Other Sources of Particulate Air Pollution on the City of Delhi for the Year 2015 & 2016 & The study analyzed Delhi's air pollution (2015), showing PM$_{2.5}$ at 295 µg/m³, PM$_{10}$ at 470 µg/m³, and AQI peaking at 435. Road dust (34-51\%) was the largest source, while vehicles contributed 12\%. Winter inversions amplified PM$_{2.5}$ by 7x. &\cite{Gopalaswami2016}
\\
\\      
 Air pollution and public health: the challenges for Delhi, India & 2017 & The study reviewed Delhi’s air pollution impact (1986–2017), linking PM$_{2.5}$, PM$_{10}$, NO$_x$, SO$_2$, and CO to COPD, asthma, and cardiovascular diseases. It highlighted data inconsistencies, monitoring gaps, and the need for standardized research on health effects. &\cite{Sharma2018}
\\
\\
Managing future air quality in megacities: A case study for Delhi& 2017	&The study showed that improving air quality in Delhi requires a combination of local emission controls and regional cooperation, targeting sources like vehicular emissions, biomass burning, and industrial pollution for effective long-term solutions &\cite{Amann2017}
\\
\\
 Increasing Potential for Air Pollution over Megacity New Delhi: A Study Based on 2016 Diwali Episode & 2018 & The study showed that PM$_{2.5}$ exceeded NAAQS on 85\% of days in Delhi, with 95\% exceedance in winter. Diwali 2016 saw a severe pollution spike due to local emissions and long-range transport from biomass burning in northwest India. Fire counts in these regions increased by ~25\% per year since 2000. Aerosol Optical Depth (AOD) rose by ~54\% since 2000, indicating increasing background pollution. &\cite{Mukherjee2018}
\\
\\        
 A Novel Approach to Understanding Delhi’s Complex Air Pollution Problem& 2019&The study analyzed Delhi’s air pollution (2019), highlighting PM$_{10}$ exceedance since 2009 and vehicular emissions rising to 72\%. Heavy commercial vehicles were the main PM source, while policies like odd-even had minimal impact. &\cite{Chakraborty2019}
 \\
 \\
 Air pollution in Delhi: biomass energy and suitable environmental policies are sustainable pathways for health safety& 2019&The study highlights that air pollution in Delhi, primarily caused by crop residue burning, vehicular emissions, and industrial activities. &\cite{Tripathi2019}
\\
\\
Who Is Responsible for Delhi Air Pollution? Indian Newspapers’ Framing of Causes and Solutions&2019&The study examines how the Indian print news media has framed the issue of Delhi air pollution, and framed responsibilities for its causes and solutions.&\cite{Bhalla2019}
\\
\\
 Four-year assessment of ambient particulate matter and trace gases in the Delhi-NCR region of India& 2020&The study analyzed air pollution trends in Delhi-NCR (2014–2017), assessing PM$_{2.5}$, PM$_{10}$, NO$_x$, SO$_2$, CO, and O$_3$ across 12 monitoring stations. Winter had the highest pollution levels, monsoon the lowest. PM$_{2.5}$ was dominated by local sources, with high spatial divergence between Delhi and NCR sites. &\cite{Hama2020}
\\
\\
 Assessing the Trend and Status of Air Quality in NCT Delhi& 2021&The study analyzed Delhi’s air quality (1991–2017), finding PM$_{10}$, NO$_2$, CO, and SPM exceeded NAAQS limits, while SO$_2$ remained within safe levels. &\cite{Kumar2021}
\\
\\
 Policy Interventions and Their Impact on Air Quality in Delhi City — an Analysis of 17 Years of Data& 2021&The study analyzed Delhi’s air pollution trends (2003–2019) and policy impacts. RSPM increased (0.98–3.19\% annually), highest at industrial sites. SO$_2$ declined (1.49–4.09\% annually), while NO$_2$ rose (5.21–6.07\% annually). &\cite{Gulia2021}
\\
\\
 Analysis of Air Pollution Data in India between 2015 and 2019& 2022&PM$_{2.5}$ and PM$_{10}$ levels exceeded national ambient air quality standards (NAAQS) by 150\% and 100\% in northern India, and by 50\% and 40\% in southern India (2015–2019). SO$_{2}$, NO$_{2}$, and O$_{3}$ mostly met NAAQS. Northern India had 10–130\% higher pollution than southern India, except for SO$_{2}$, which remained similar. No significant pollution trend was observed over five years. &\cite{Sharma2022}
\\
\\
 Trends and Variability of PM$_{2.5}$ at Different Time Scales over Delhi: Long-term Analysis 2007–2021& 2022&The study analyzed PM$_{2.5}$ trends in Delhi (2007-2021) and found a decline, but levels remained above safe limits, with minimal impact from policy measures.  &\cite{Chetna2023}
\\
\\
 Long-term trend analysis of criteria pollutants in megacity of Delhi: Failure or success of control policies& 2022&The study analyzed air pollutants in Delhi (2000--2019) and found PM$_{2.5}$ and NO$_{2}$ declined, while SO$_{2}$ and CO rose, with transport being the main source of NO$_{x}$ and PM$_{2.5}$, and power plants contributing to SO$_{2}$. &\cite{Verma2022}
\\
\\
 Variation in Air Quality over Delhi Region: A Comparative Study for 2019 and 2020& 2022&The study analyzed air quality variations in Delhi (2019–2020), comparing pollutant levels before, during, and after COVID-19 lockdowns, showing a significant reduction in PM$_{2.5}$ (~11.6\%), NO$_{x}$ (~7\%), SO2 (~3.7\%), CO (~20.7\%), and ozone (~7.7\%) during lockdown, with levels rising post-lockdown. &\cite{Shankar2022}
\\
\\
 What Is Polluting Delhi’s Air? A Review from 1990 to 2022& 2023&1990–2022 study analyzed vehicular emissions, road and construction dust, biomass burning, industries, and long-range transport. PM$_{2.5}$ in 2021–22 was 100 µg/m³, 20× WHO limit. Pollution improved ~30\% since pre-CNG era but remains severe. &\cite{Guttikunda2023}
\\
\\
 Decoding temporal patterns and trends of PM$_{10}$ pollution over Delhi: a multi-year analysis (2015–2022)& 2024&The study analyzed PM$_{10}$ trends in Delhi (2015–2022) using data from 37 stations. PM$_{10}$ declined (-7.57 µg/m³ per year ambient, -8.45 µg/m³ de-seasonalized). &\cite{Chetna2024}
\\
\\
 Long-term meteorology-adjusted and unadjusted trends of PM$_{2.5}$ using the AirGAM model over Delhi, 2007–2022& 2024&The study analyzed PM$_{2.5}$ trends in Delhi (2007–2022) using the AirGAM model, adjusting for meteorological factors. PM$_{2.5}$ declined by 14 µg/m³ (unadjusted) and 18 µg/m³ (meteorology-adjusted) over the period. &\cite{Chetna2024b}
\\
\\
 Does Stubble Burning Really Contribute in Delhi’s Air Pollution? Evidences from Ground, Model, and Satellite Data& 2024&The study analyzed stubble burning's impact on Delhi’s air pollution (2019–2020) using ground observations, geophysical models, and satellite data.&\cite{Kundu2024}
\\
\\
 Insights into the Air Quality Indices and its Linkage with Diwali Festival Celebrations in Delhi, India in November 2023: A Case Study& 2024&The study analyzed air quality indices in Delhi during Diwali (November 2023), highlighting a significant spike in pollution levels due to festival-related emissions.&\cite{Bhola2024}
\\
\\
Critical Analysis of PM$_{2.5}$ in Delhi Region to Strategize Effective Air Pollution Management Plan&2024&The study assesses the status of a critical pollutant (PM$_{2.5}$) at selected locations in Delhi city and the subsequent formulation of source-specific control measures considering the surrounding local activities.&\cite{Jha2024}
\\
\\
Air pollution and its effects on emergency room visits in tertiary respiratory care centers in Delhi, India	& 2024 &The study finds that increased air pollution in Delhi significantly correlates with a rise in emergency room visits for acute respiratory symptoms, particularly due to high levels of particulate matter and nitrogen dioxide.&\cite{Kumar2024}
\\
\\
 Critical review of air pollution contribution in Delhi due to paddy stubble burning in North Indian States& 2025&The study reviewed stubble burning’s impact on Delhi’s air quality (2015–present), analyzing ground-based, satellite, and modeling data. It assessed biomass burning trends across seasons, highlighting uncertainties in contribution estimates.&\cite{Goyal2025}
\\
\\
\hline
\\
    \caption{Summary of Literature on air pollution in Delhi}
    \label{tab:literature_review}
\end{longtable}
\end{scriptsize}

\newpage

The reviewed literature as shown in \autoref{tab:literature_review}  presents: 
\begin{itemize}
    \item 7 papers -- Long-term trends and analysis of air pollution in Delhi.
    \item 4 papers -- Source-specific contributions to air pollution in Delhi.
    \item 4 papers -- Impact of seasonal and meteorological factors.
    \item  4 papers -- Effectiveness of control policies and interventions.
    \item  4 papers -- Impact of stubble burning on air pollution in Delhi.
    \item 2 papers -- PM$_{2.5}$ and PM$_{10}$ temporal trends.
    \item  1 papers -- Media and data analysis of air pollution in Delhi.
\end{itemize}

Most of the previous studies focused on historical trends without integrating real-time analysis. Some studies have analyzed pollution seasonally, but have not exhibited its yearly impact. Studies have provided limited data over the years on probable causes of pollution. The present study focuses on analyzing trends on four probable contributors to air pollution in Delhi over 10 years. The multiyear trend in each case provides a deeper understanding of how significantly each contributes to air pollution in Delhi. The study also provides a correlation of these factors with the multi-year air quality trend. 

\section{Research Questions}\label{sec3}
To analyze the Delhi air pollution crisis over a decade, the study addresses the following questions:\\
\\
RQ1- What are the decadal levels of AQI in Delhi?\\
RQ2- How have the probable contributors to Delhi air pollution evolved over the years?\\
RQ3- What is the correlation of probable contributors with the AQI?

\section{Methodology}\label{sec4}

\begin{figure}[H]
    \centering
    \includegraphics[width=0.75\linewidth]{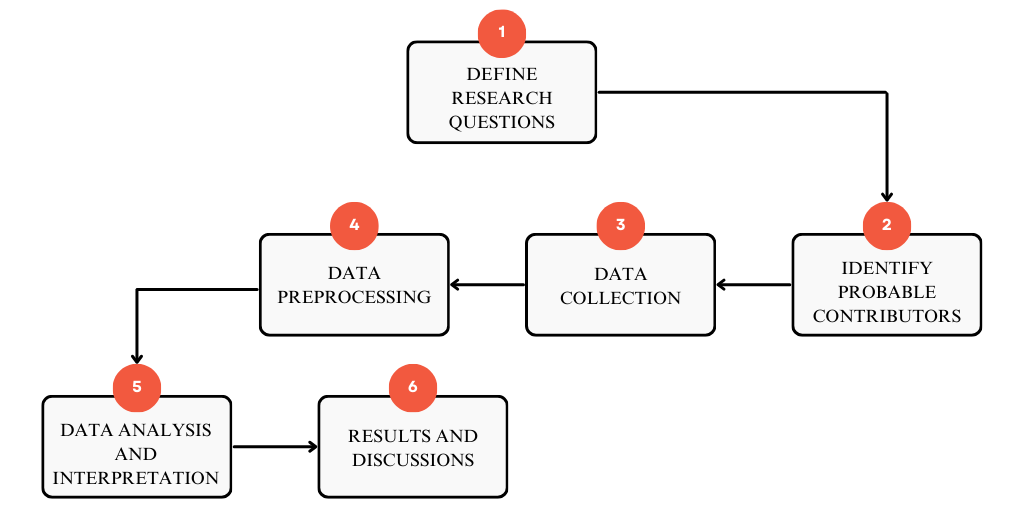}
    \caption{\centering Stepwise methodology followed for the study}
    \label{fig:2}
\end{figure}
The overall stepwise methodology followed for the study is presented in \autoref{fig:2} . Based on the key pollutants causing air pollution, probable contributors were selected. This was followed by collecting data for these contributors over the years from various publicly available sources. The study later discusses data collection in detail. Using the dataset for each contributor, its trend over the years was analyzed using statistical methods such as bar charts with trendlines.

\begin{figure}[H]
    \centering
    \includegraphics[width=0.75\linewidth]{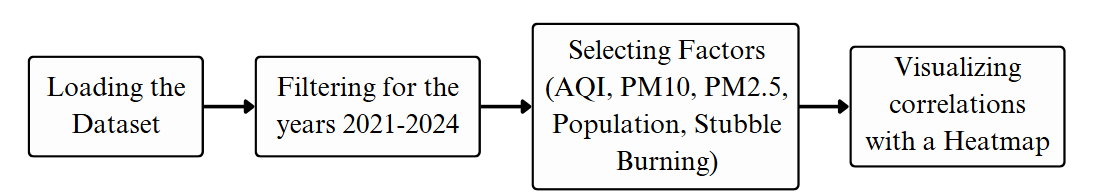}
    \caption{\centering Stepwise methodology followed for correlation analysis}
    \label{fig:corr}
\end{figure}

Moreover, the correlation between AQI and various factors such as PM$_{2.5}$, PM$_{10}$, population, and stubble burning was visualized using heatmaps -- a colored chart used to visually represent the correlation between different factors. The stepwise methodology followed is presented in \autoref{fig:corr} . This was done using Python \cite{Python2025} -- a computer programming language with ready-made tools (libraries) used to process and analyze data. 
The tools used to generate heatmaps are as follows.
\begin{itemize}
    \item Pandas \cite{Pandas2025} -- it is a library in Python that is used to load and process data.
    \item Seaborn \cite{Seaborn2025} -- it is a library in Python that is used to create graphs and visualizations. In this study seaborn was used to create heatmaps. The darker color indicates a stronger relationship, while the lighter color indicates a weaker relationship between the different factors selected.
    \item Matplotlib \cite{Matplotlib2025} -- it is a library in Python that helps to display visual graphs.
\end{itemize}

Correlation: It is used to mathematically measure how two factors are related to each other. The correlation value (r) is calculated, it ranges between -1 and 1, where;
\begin{itemize}
    \item Positive correlation (+1) indicates that if one factor increases, the other increases as well.
    \item Negative correlation (-1) indicates that if one factor increases, the other decreases.
    \item No correlation (0) indicates that the two factors are not affected by each other.
\end{itemize}

Mathematical formula for the correlation value (r), also known as the Pearson correlation coefficient:

\begin{equation}
r = \frac{\sum (X_i - \bar{X})(Y_i - \bar{Y})}{\sqrt{\sum (X_i - \bar{X})^2} \cdot \sqrt{\sum (Y_i - \bar{Y})^2}}
\end{equation}

where:
\begin{itemize}
    \item \( r \) = Pearson correlation coefficient (ranges from -1 to 1)
    \item \( X_i, Y_i \) = Individual data points for factors \( X \) and \( Y \)
    \item \( \bar{X}, \bar{Y} \) = Mean (average) of the factors \( X \) and \( Y \)
    \item \( \sum \) = Summation symbol, representing the sum of all terms
    \item \( \sqrt{} \) = Square root function
\end{itemize}

In addition, a 10-year literature review was conducted using the following tools.
 \begin{itemize}
     \item IEEE Xplore
     \item SpringerLink
     \item ScienceDirect
     \item Google Scholar
 \end{itemize}
The search keywords used were 'Delhi' AND 'trends' AND 'air pollution'.

In general, the study followed an evidence-based discussion and a multifaceted approach to ensure a comprehensive assessment of Delhi's air pollution.
\section{Data Collection}\label{sec5}

Data for this study were collected from various publicly available sources, such as government websites and open access datasets. The primary sources of data include the following.

\begin{itemize}
    \item The air quality data and the stubble burning data were obtained from the Press Information Bureau (PIB) \cite{PIB2025Link} of the Government of India and the reports published by the Central Pollution Control Board (CPCB) \cite{CPCB2025}. PIB provided the average monthly AQI for each year from the year 2018 to 2024.
    \item Vehicular data were sourced from the OpenCity data portal \cite{OpenCity2025}.
    \item GHG emissions were sourced from the environment department of the Delhi government \cite{DelhiEnv2025}.
    \item EV registrations and Delhi metro ridership data were sourced from the Open Government Data (OGD) Platform India \cite{OGD2025}.
    \item The population growth trends for Delhi were sourced from the World Population Review database \cite{WorldPop2025}.  
\end{itemize}

\section{Results and Discussions}\label{sec6}

\subsection{Air Quality Analysis - RQ1}\label{subsec6.1}
Despite various mitigation efforts, the Delhi Air Quality Index (AQI) continues to exhibit persistent levels of air pollution.

Delhi has identified 13 hotspots, namely Narela, Bawana, Mundka, Wazirpur, Rohini, R.K. Puram, Okhla, Jahangirpuri, Anand Vihar, Punjabi Bagh, Mayapuri, Dwarka, in consultation with EPCA and CPCB, as the said spots have higher levels of air pollution compared to other areas. Hotspots have been identified depending on the annual average of PM$_{10}$ and for PM$_{2.5}$ more than 100 $\mu$g/m$^{3}$
. This is based on data from nearby continuous air quality monitoring stations (within radius of 2km) established by Delhi Pollution Control Committee (DPCC). \cite{DelhiHotspots}

\begin{figure}[H]
    \centering
    \includegraphics[width=0.5\linewidth]{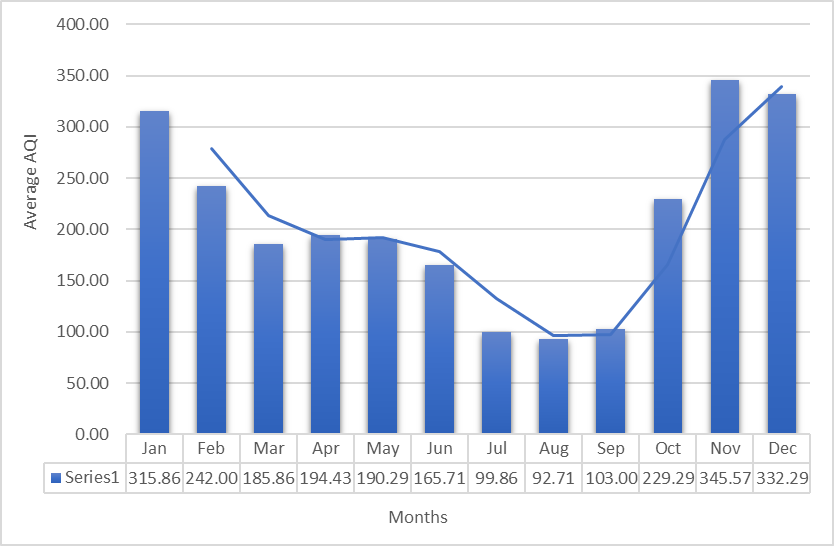}
    \caption{\centering Average Monthly AQI | FY 2018-2024}
    \label{fig:3}
\end{figure}

The PIB (2024, December 31) accounted still/very low- speed wind conditions for an abnormally high average AQI of 355 for the month of January. For the months of April-June, poor AQI was reasoned to be due to exceptionally prolonged dry spells coupled with high-speed winds leading to regional transport of fine dust and PM from the adjoining areas as well as transboundary.  The month of May had suffered the worst ever AQI average in 2024, on this account for the period between 2018 and 2024. \cite{PIB2025b}
\\
\autoref{tab:2} shows average monthly AQI levels from the year 2018 to 2024 and \autoref{fig:3} shows the average AQI for each month calculated from the year 2018 to 2024. November has the highest average of 345.57 and August has the lowest average of 92.71.

\begin{table}[htbp]
\noindent
\centering
\begin{tabular}{|c|c|c|c|c|c|c|c|}
\hline
\textbf{Month} & \textbf{2018} & \textbf{2019} & \textbf{2020} & \textbf{2021} & \textbf{2022} & \textbf{2023} & \textbf{2024} \\
\hline
Jan & 328 & 328 & 286 & 324 & 279 & 311 & 355 \\
Feb & 243 & 242 & 241 & 288 & 225 & 237 & 218 \\
Mar & 203 & 184 & 128 & 223 & 217 & 170 & 176 \\
Apr & 222 & 211 & 110 & 202 & 255 & 179 & 182 \\
May & 217 & 221 & 144 & 144 & 212 & 171 & 223 \\
Jun & 202 & 189 & 123 & 147 & 190 & 130 & 179 \\
Jul & 104 & 134 & 84 & 110 & 87 & 84 & 96 \\
Aug & 111 & 86 & 64 & 107 & 93 & 116 & 72 \\
Sep & 112 & 98 & 116 & 78 & 104 & 108 & 105 \\
Oct & 269 & 234 & 266 & 173 & 210 & 219 & 234 \\
Nov & 335 & 312 & 328 & 377 & 320 & 373 & 374 \\
Dec & 360 & 337 & 332 & 336 & 319 & 348 & 294 \\
\hline
\end{tabular}
\caption{\centering Yearly Monthly AQI Data (2018-2024)}
\label{tab:2}
\end{table}

\subsection{Analysis of Probable Contributors - RQ2 }\label{subsec6.2}
Various sources, including stubble burning cases in neighboring cities, vehicular emissions, industrial activities, population growth, and impact after mitigation strategies were analyzed. 

\subsubsection{Stubble Burning}\label{subsubsec6.2.1}
Stubble burning is the practice of intentionally setting fire to the straw stubble that remains after grains, such as rice and wheat, have been harvested. The technique is used to quickly and cheaply clear fields. Crop Residue Burning (CRB) is often associated as a huge contributor of fine particulates (e.g. PM$_{2.5}$) in the Delhi-NCR (National Capital Region) during the winter season every year. These fine particles (PM$_{2.5}$) pose a higher health risk to the public, especially the farmers living in CRB and nearby areas \cite{CPCB}. Stubble burning releases several particulate precursors that increase the levels of organic substances capable, and fine particles released can trap the toxic heavy metals and gases due to their adsorptive surface which intensify their adverse health impacts. 

\begin{figure}[H]
    \centering
    \includegraphics[width=0.5\linewidth]{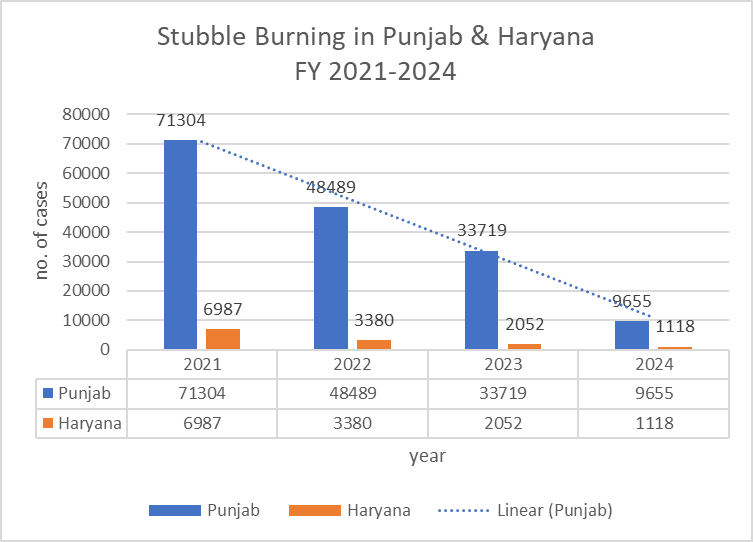}
    \caption{\centering Stubble Burning in Punjab and Haryana FY: 2021-2024 }
    \label{fig:4}
\end{figure}

The present study shows the trends for the years 2021 to 2024 for the neighboring states of Punjab and Haryana, which can be referred from \autoref{fig:4}. For Punjab, the number of paddy stubble burning incidents has consistently decreased: 71,304 (year 2021), 48,489 (year 2022), 33,719 (year 2023), and 9,655 (year 2024). Similarly, a decline can be observed for Haryana: 6,987 (year 2021), 3,380 (year 2022), 2,052 (year 2023) and 1,118 (year 2024)\cite{PIB2025}.

Various measures taken by the government contributed to this decline, including: (i) The government has subsidized the distribution of machinery such as Happy Seeders, Super Straw Management System (SMS) and rotavators to facilitate in situ stubble management. More than 117,672 machines in Punjab, 80,071 in Haryana, and 7,986 in Uttar Pradesh-NCR (as of October 2023). (ii) Rs. 3,333 crores are allocated under the Crop Residue Management Scheme for Punjab, Haryana, Uttar Pradesh, and Delhi. (iii) In Haryana, 22 complaints led to 16 arrests, investigations against 100 farmers, and 300 fines issued.
Despite all the corrective measures and the steep decline, the air quality index does not appear to be affected \cite{PIB2025}.

\subsubsection{Vehicular Emissions}\label{subsubsec6.2.2}

Vehicular emissions are the harmful by-products created when gasoline and diesel are burned in vehicles. Vehicular emissions include harmful gases such as Carbon Dioxide (CO$_{2}$), NO$_{x}$, CO, Volatile Organic Compounds (VOCs) and PM. 

The Center for Science and Environment (CSE), a Delhi-based research and advocacy group, said that vehicles in the capital contributed more to air pollution than stubble burning in neighboring states during the period from October 12 to November 3, 2024 \cite{Hindu2025}. In addition, it further added that slow incremental change in public transport systems, lack of integration, inefficient last mile connectivity, and hidden subsidy for the use of personal vehicles cannot address this mobility crisis in the city. 

The neighboring districts accounted for 34.97\% of the city’s air pollution, while the contribution of local sources of Delhi was 30.34\%. The contribution of farm fires was only 8.19\% during the period, while the remaining sources of pollution were in areas beyond the NCR. Vehicular emissions contributed the most (51.5\%) while the share of dust particles was 3.7\%. 

\begin{figure}[H]
    \centering
    \includegraphics[width=1\linewidth]{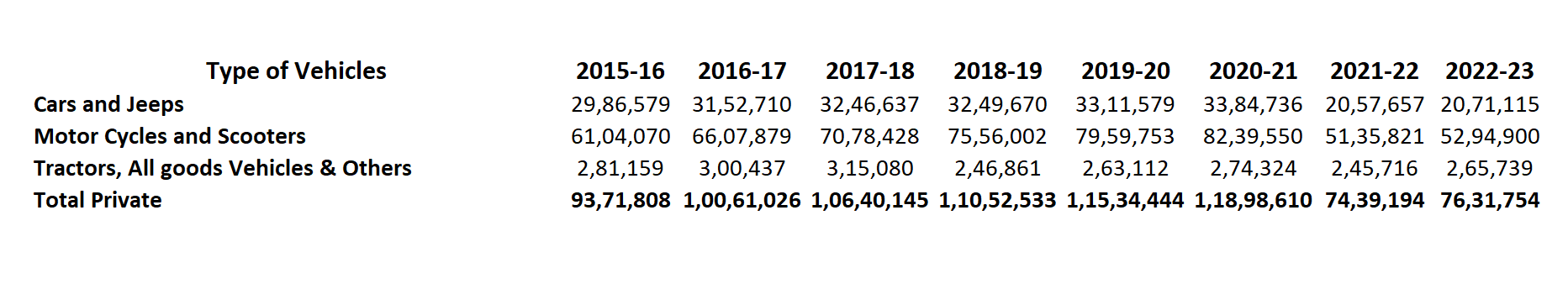}
    \caption{\centering Private Vehicle Type - Wise Data FY: 2015-2023}
    \label{fig:5}
\end{figure}

\begin{figure}[H]
    \centering
    \includegraphics[width=0.5\linewidth]{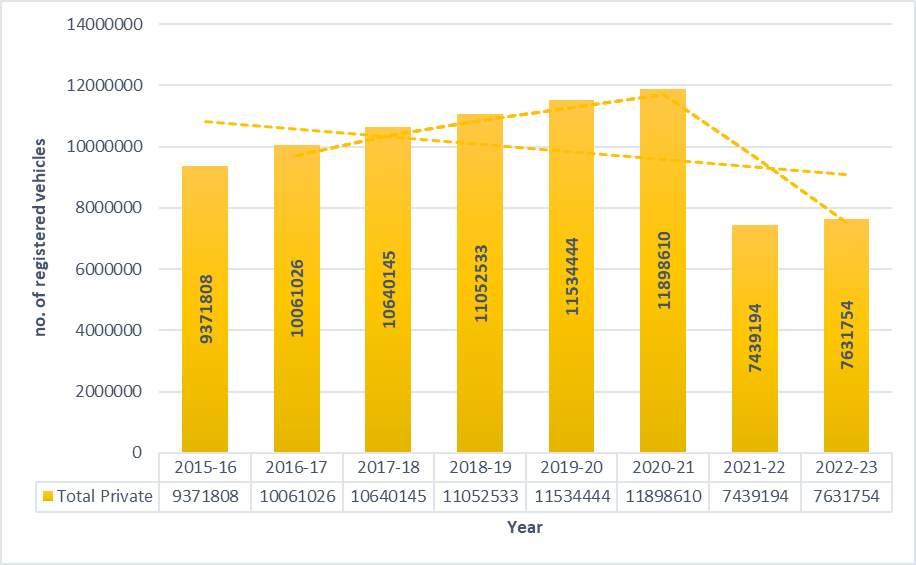}
    \caption{\centering Private Vehicles Registered in Delhi FY : 2015-2023}
    \label{fig:6}
\end{figure}

\begin{figure}[H]
    \centering
    \includegraphics[width=1\linewidth]{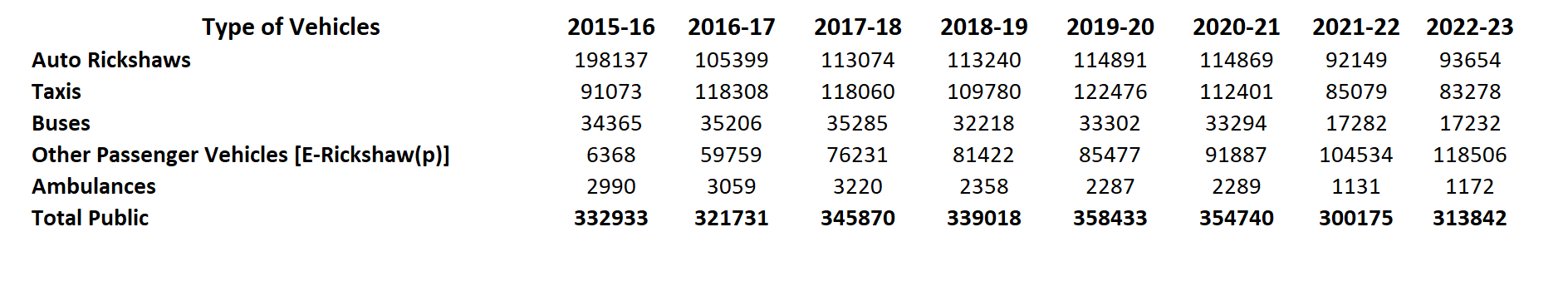}
    \caption{Public Vehicle Type - Wise Data FY: 2015-2023}
    \label{fig:7}
\end{figure}

\begin{figure}[H]
    \centering
    \includegraphics[width=0.5\linewidth]{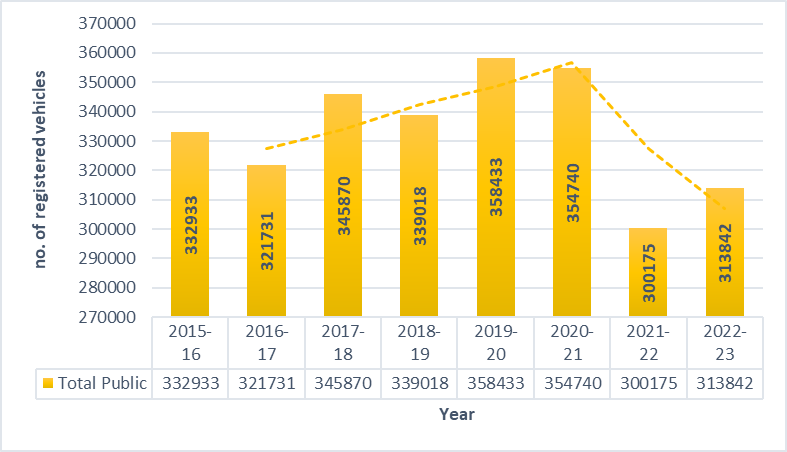}
    \caption{\centering Public Vehicles Registered in Delhi FY : 2015-2023}
    \label{fig:8}
\end{figure}

To analyze trends in vehicular emissions over the years, the number of registered vehicles in Delhi from the year 2015 to 2023 was studied -- presented in \autoref{fig:5}, \autoref{fig:6}, \autoref{fig:7}, and \autoref{fig:8}. Public vehicles (Auto Rickshaws, Taxis, Buses, Other passenger vehicles [E-Rickshaw(p), Ambulances]) registered were as follows: 332,933 (2015-16), 321,731 (2016-17), 345,870 (2017-18), 339,018 (2018-19), 358,433 (2019-20), 354,740 (2020-21), 300,175 (2021-22) and 313,842 (2022-23). The private vehicles (Cars and Jeeps, Motorcycles and Scooters, Tractors, All goods vehicles, and Others) registered were: 9,371,808 (2015-16), 10,061,026 (2016-17), 10,640,145 (2017-18), 11,052,533 (2018-19), 11,534,444 (2019-20), 11,898,610 (2020-21), 7,439,194 (2021-22) and 7,631,754 (2022-23). \cite{OpenCityData}

\subsubsection{Industrial Activities}\label{subsubsec6.2.3}
Delhi falls under one of the main industrial regions in India, i.e., the Gurgaon-Delhi-Meerut industrial region, which is one of the most significant economic regions of the nation and in the National Capital Territory (NCT) of Delhi, industrial areas cover 51.81 km${^2}$ of the landscape \cite{Parveen2021}. The number of working factories in Delhi increased from 8,219 in 2011 to 8,690 in 2022. The maximum number of factories in Delhi is registered in three major industrial groups (i) textiles product, (ii) Basic metal \& alloy, followed by (iii) Metal products and Parts machinery. \cite{DelhiGov}
\noindent
\begin{figure}[H]
    \centering
    \includegraphics[width=0.5\linewidth]{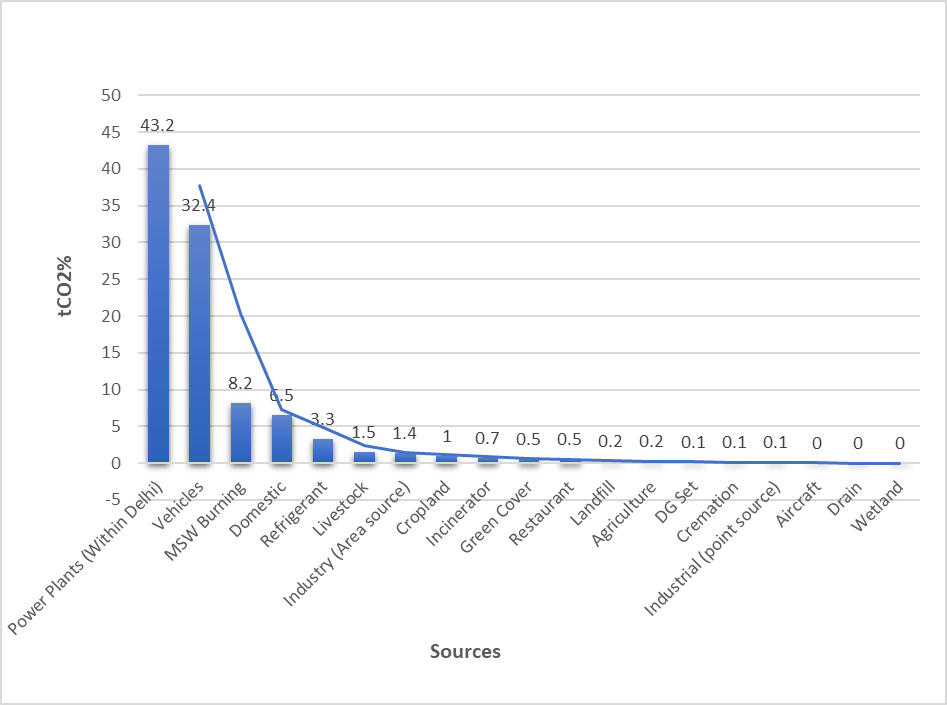}
    \caption{\centering GHG Emissions | FY 2014}
    \label{fig:9}
\end{figure}

\autoref{fig:9} shows trend of data sourced from a project on a comprehensive study of GHG in Delhi (year 2014), which showed that the total carbon equivalent (tCO$_{2}$e) GHG emission load in Delhi was estimated at 37.91 million tonnes in 2014. The top four contributors to tCO$_{2}$e emission were:
\begin{itemize}
    \item Power plants (electricity generation and consumption) -- 43.2\%
    \item Vehicles -- 32.4\%
    \item Burning of Municipal Solid Waste (MSW) -- 8.2\%
    \item Domestic fuel -- 6.5\%
\end{itemize}

\begin{figure}[H]
    \centering
    \includegraphics[width=0.5\linewidth]{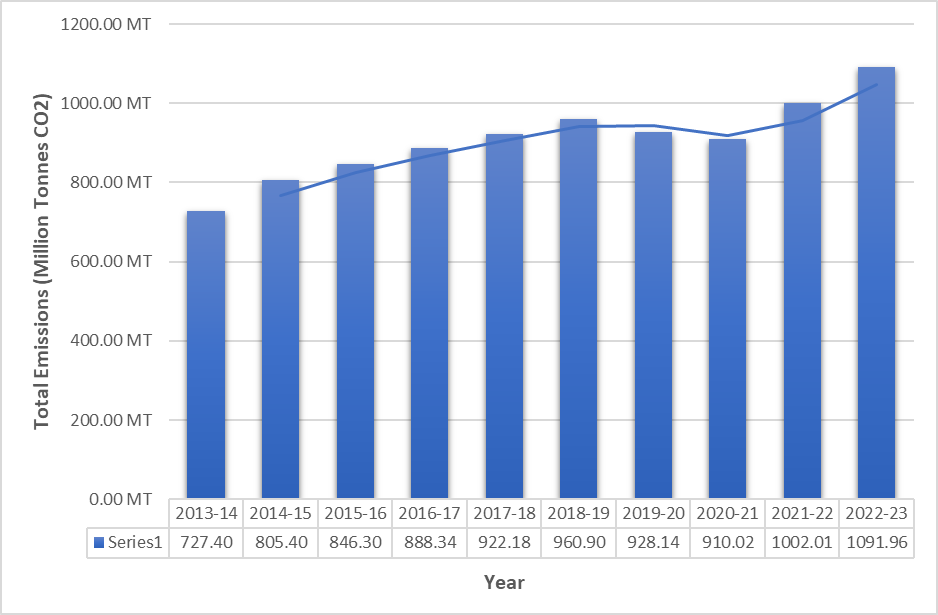}
    \caption{\centering Total CO2 Emissions in Million Tonnes | FY 2013-2023}
    \label{fig:10}
\end{figure}

The comprehensive data studies various sources and their contribution to Delhi air pollution. The contribution to tCO$_{2}$e emission for cropland and agriculture was 1\% and 0.2\%, respectively \cite{DelhiEnvDept}. Throughout the years, these numbers have only increased. The total emission of the power sector for 2018-19 to 2022-23, in million tonnes of CO$_{2}$ showed a drastic increase from 960.90 (FY 2018-2019) to 1091.96 (FY 2022-2023). It shows that cross-border electricity transfers and renewable energy contributions affect emissions \cite{CEA2024}. \autoref{fig:10} presents trend of total CO$_{2}$ emissions in million tonnes from the year 2013 to 2023. The highest emissions for FY 2022-23 and the lowest FY 2013-14.

After analyzing the contribution of fossil fuel power plants to pollution (FY 2022-2023), it was seen that Lignite (a fossil fuel that contains carbon. When lignite is burned, it releases carbon dioxide (CO$_{2}$), methane (CH$_{4}$), nitrous oxide (N$_{2}$O) and other pollutants) is the highest emitter, producing 1.286 tCO$_{2}$/MWh (Megawatt-hour); Coal follows with 0.978 tCO$_{2}$ / MWh, Gas stations produce 0.478 tCO$_{2}$ / MWh.

The data show that renewable energy generation has also increased from 53.06 BU in 2013-14 to 203.55 BU in 2022-23. The average CO$_{2}$ emission factor has decreased from 0.774 tCO2/MWh in 2013-14 to 0.716 tCO$_{2}$ / MWh in 2022-23. This suggests that renewables are helping mitigate CO$_{2}$ emissions per unit of electricity generated. 

However, as we saw earlier, total emissions each year continue to rise, which means that, while per-unit emissions have improved, overall industrial and energy demands have outpaced the benefits of renewable energy growth. 

\subsubsection{Population Growth}\label{subsubsec6.2.4}
Delhi is the fifth most populous city in the world and the largest city in India in area \cite{WorldPopReview}.
Human activities that burn fossil fuels and biomass, including transportation, cooking and heating, power generation, waste incineration, and industry, lead to the release of dangerous air pollutants.

A study on 'Population density and urban air quality' concluded that air quality as measured by the aggregate AQI decreases with population density, with an elasticity of 0.14 \cite{Borck2020}. Delhi’s population is growing at the rate of 2.63\%. As urbanization increases, so does the consumption of resources. According to The Wire (16 February/2023), households are now spending 25\% more on these essentials compared to 2021 \cite{Wire2025}.

\begin{figure}[H]
    \centering
    \includegraphics[width=0.5\linewidth]{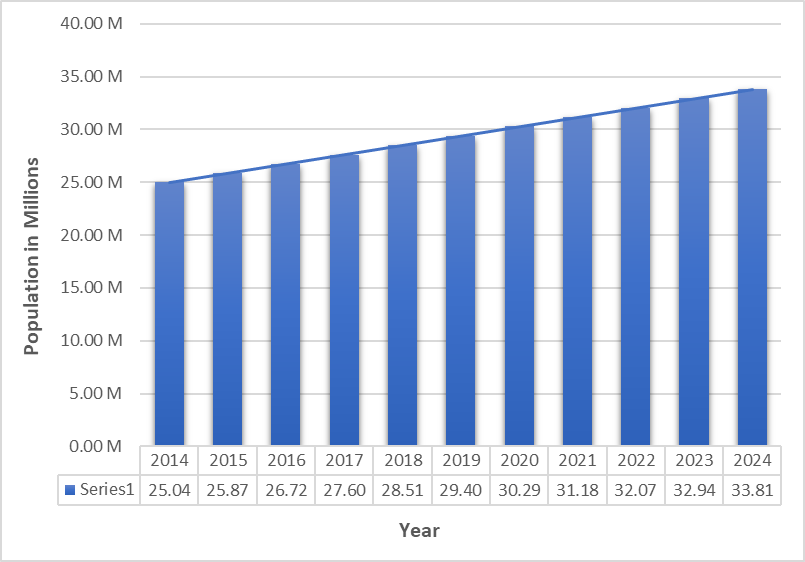}
    \caption{\centering Delhi's Population Growth | FY 2014-2024}
    \label{fig:11}
\end{figure}

The population of Delhi from the year 2014 to 2024 is shown in \autoref{tab:3}. The corresponding visualization is present in \autoref{fig:11}
\begin{table}[htbp]
\noindent
\centering

\begin{tabular}{|l |l |l|} \hline  
\textbf{Year} & \textbf{Population}& \textbf{In Millions}\\ \hline  
2014 & 2,50,39,100 & 25.04 M \\ \hline  
2015 & 2,58,65,900 & 25.87 M \\ \hline  
2016 & 2,67,20,000 & 26.72 M \\ \hline  
2017 & 2,76,02,300 & 27.60 M \\ \hline  
2018 & 2,85,13,700 & 28.51 M \\ \hline  
2019 & 2,93,99,100 & 29.40 M \\ \hline  
2020 & 3,02,90,900 & 30.29 M \\ \hline  
2021 & 3,11,81,400 & 31.18 M \\ \hline  
2022 & 3,20,65,800 & 32.07 M \\ \hline  
2023 & 3,29,41,300 & 32.94 M \\ \hline  
2024 & 3,38,07,400 & 33.81 M \\ \hline 

\end{tabular}
\caption{\centering Delhi Population (2014-2024)}
\label{tab:3}
\end{table}

\subsubsection{Impact After Mitigation Strategies}\label{subsubsec6.2.5}
In recent years, various mitigation strategies have been implemented to combat air pollution in Delhi. Starting with the Graded Response Action Plan (GRAP), it is implemented as an emergency response system with four stages based on the AQI. Stage I (poor, AQI 201-300): (i) Dust mitigation measures at Construction and Demolition (C\&D) sites. (ii) Ban on sets of Diesel Generator (DG) running on pure diesel. Stage II (Very Poor, AQI 301-400): (i) Stricter enforcement of dust control at C\&D sites. (ii) Ban on DG sets running purely on diesel. Stage III (Severe, AQI 401-450): (i) Ban on C\&D activities in NCR. Restrictions on BS III petrol and BS IV diesel light motor vehicles (LMV). (ii) Stage IV (Severe+, AQI \texttt{>} 450): Ban on entry of truck traffic into Delhi (except essential services). (iii) Restrictions on diesel-run medium and heavy goods vehicles. Ban on activities of linear public projects such as road construction. \cite{DPCC}
To control construction and demolition pollution, the following measures were taken: A web portal was built to monitor dust control measures at construction sites, ban of roadside debris storage, mandate to recycle construction and demolition waste.\cite{DPCC}

The mitigation of industrial pollution includes the following: (i) Shift to cleaner fuels such as PNG, electricity, and LPG. (ii) Ban on polluting fuels such as coal, furnace oil, and pet coke. (iii) Mandatory emission control technologies for permitted fuels. (iv) 1,727 industries converted to PNG.
More mitigation strategies imposed by the Delhi government include: (i) Ban and regulation of diesel generators. (ii) Ban on burning garbage, biomass and waste plastics. (iii) Fines imposed for violations under orders of the National Green Tribunal (NGT). (iv) Anti-smog guns at major construction sites. (v) Smog towers installed in high-pollution zones. Regular sprinkling of water on the roads. (vi) Monitoring of road dust pollution through a dedicated task force. \cite{GlobalRailway2025}

The Delhi Metro Rail Corporation (DMRC) has been certified by the United Nations (UN) as the first metro rail and rail-based system in the world to get carbon credits for reducing GHG emissions.  \cite{DataGov}. The number of daily average ridership in Delhi metro from 2018 to 2022 is present in \autoref{tab:4}.

\begin{table}[h]
\centering

\begin{tabular}{|c|c|} \hline   
\textbf{Year}& \textbf{No. of Daily Avg Ridership}\\ \hline 
2018-19 & 45.44 L \\ \hline 
2019-20 & 50.65 L \\ \hline 
2020-21 & 17.10 L \\ \hline 
2021-22 & 24.77 L \\ \hline 
Jun-22 & 41.21 L \\ \hline

\end{tabular}
\caption{Delhi Metro | Daily Average Ridership}
\label{tab:4}
\end{table}
It is observed that the daily average ridership of Delhi Metro has remained inconsistent over the years. Research has consistently shown that EVs produce fewer greenhouse gases and air pollutants over their lifetime compared to petrol and diesel cars—even when factoring in vehicle production and electricity generation.
Delhi EV policy stated that the Delhi Government aims to have 1 of every 4 vehicles sold in Delhi by 2024, to be an EV \cite{DelhiEVPolicy}.

\begin{figure}[H]
    \centering
    \includegraphics[width=0.5\linewidth]{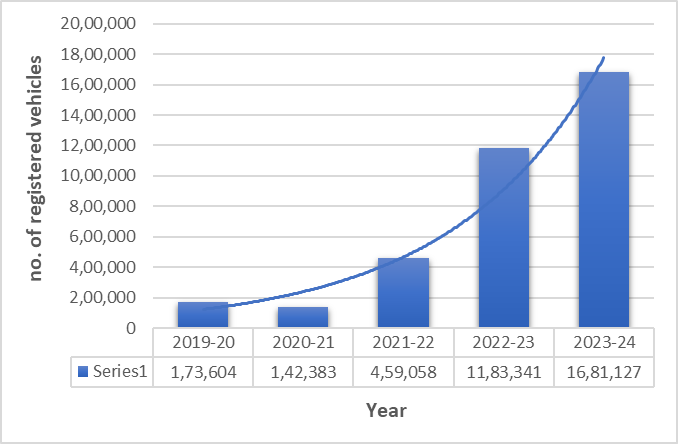}
    \caption{\centering Electric Vehicles Registered in Delhi | FY 2019-2024}
    \label{fig:12}
\end{figure}

Although the 25\% target has not yet been fully achieved, the data present in \autoref{fig:12}  show a clear exponential growth in EVs registered in Delhi from the year 2019 to 2024 \cite{DataGovEV}.

\subsection{Correlation Analysis -RQ3}\label{subsec6.3}
\begin{figure}[H]
    \centering
    \includegraphics[width=0.5\linewidth]{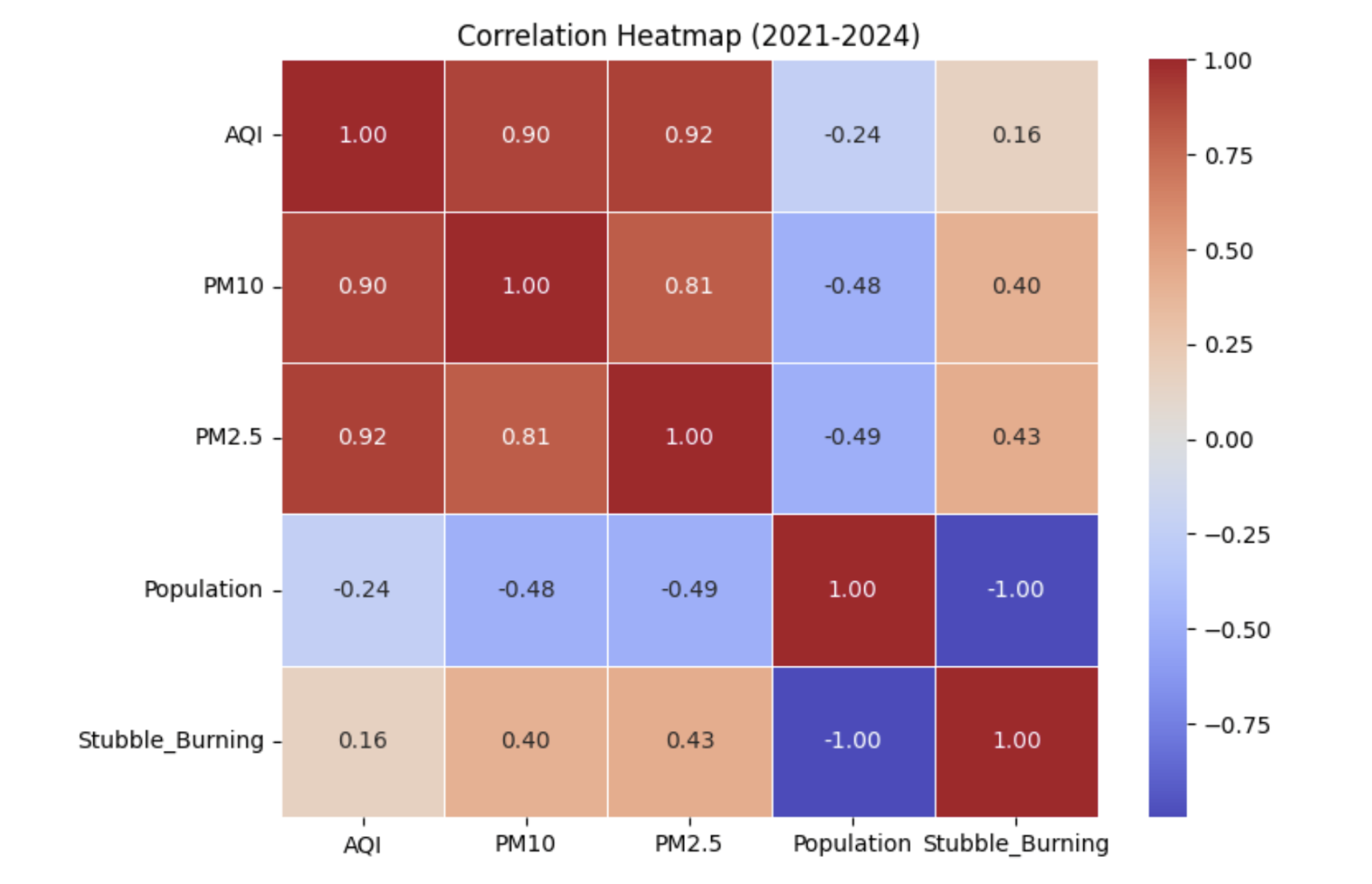}
    \caption{\centering Heatmap | AQI, PM$_{10}$, PM$_{2.5}$, Population, Stubble Burning}
    \label{fig:14}
\end{figure}

To understand the main contributors to Delhi’s air pollution, the study tried to evaluate the correlation between various contributors and the air quality index (AQI) over the years as shown in \autoref{fig:14} .

High correlations were observed between AQI and PM$_{2.5}$ and PM$_{10}$. It indicates that the contribution of particulate matter significantly influences AQI. PM$_{2.5}$ can come from sources such as vehicles, fires, wood burning, and tobacco smoke. PM$_{10}$ can come from sources such as industries, transportation (exhaust of cars and trucks), and construction.

A low correlation was observed between AQI and stubble burning. It suggests that burning has had a limited effect on the AQI over the years. 

A negative correlation was observed between population and AQI, indicating that the growing population over the years has not had an impact on air quality.

AQI levels have not improved despite a steep decline in stubble burning cases (2021-2024). Vehicular emissions accounted for 51.5\% of Delhi’s air pollution in late 2024, surpassing stubble burning at 8.19\%. No improvement in air quality was observed even after the adoption of EVs, conversion to CNG, metro expansion, and dust control measures. The study found that industrial activities and energy demand continue to increase year by year. In addition, seasonal variations and meteorological factors affect air quality.

\section{Conclusion}\label{sec7}

Delhi’s surging air pollution cannot be attributed to a single factor. The impact of multiple mitigation efforts taken has remained negligible. The citizens of Delhi have unwillingly adapted to prolonged exposure to dangerously high levels of pollution. Delhi’s air quality has remained persistently poor in the last decade, already paving the way for long-term health impacts on every citizen of Delhi.

Poor air quality has cost far more than just the air we breathe—it has disrupted education, halted economic activities, and deteriorated public health. Clean air should no longer be considered a privilege but a basic human necessity. Seasonally worsening pollution has led to schools, colleges, and offices shutting down—yet the response remains insufficient. The long-term health effects of this crisis are nothing short of poisonous and will impact generations to come.

The improvement of air quality in Delhi must be treated as an urgent priority, not as an afterthought. People must recognize that pollution is not just an inconvenience—it is a life-threatening emergency. Rather than forcing our lungs to adapt to this toxic air, we must demand accountability, stricter regulations, and sustainable solutions.

The study captures the alarming state of Delhi’s air quality as it stands today. The data used is drawn directly from government records and recent reports, making the findings credible and highly transparent. 
The study shows that Delhi’s air pollution is a multifaceted crisis, a mixture of vehicular emissions, industrial growth, seasonal effects, and weak enforcement over a unique time frame that has not been explored before. 

Existing strategies like GRAP and the Odd-Even scheme have shown little to no improvement. It is time to shift from temporary fixes to long-term and impactful reforms. Delhi does not need adaptation—it needs transformation.

\section*{Declarations}

\begin{itemize}
\item Funding - Not applicable.
\item Conflict of interest/Competing interests - The authors declare no competing interests. 
\item Ethics approval and consent to participate - The authors declare no ethical violation during the preparation of this manuscript. 
\item Consent for publication - The authors approves publication in this journal.  
\item Data availability - Datasets used are available.
\item Materials availability - Not applicable.
\item Code availability - Not applicable.
\item Author contribution - All authors contributed equally in this study.
\end{itemize}

\setlength{\bibsep}{0pt}
\renewcommand{\bibfont}{\scriptsize} 


\end{document}